\documentclass[a4paper,11pt]{article}
\usepackage[a4paper,left=3cm,right=3cm,top=3cm,bottom=2.5cm]{geometry}
\usepackage[utf8]{inputenc}
\usepackage{amsmath,amssymb}
\usepackage{graphicx}
\usepackage{color}

\title{Connecting Neural Network Training Dynamics with Effective Field Theories}
\author{Sven Krippendorf, Michael Spannowsky}

\begin{document}

\begin{flushright}
LMU-ASC 08/22\\
IPPP/22/04 \\
 \today\\[2cm]
 \end{flushright}
\begin{center}
{\LARGE\bfseries A duality connecting neural network\\ and cosmological dynamics}\\[5mm]
\vspace{1cm}

{\bf Sven~Krippendorf}${}^{\,a,}$\footnote{sven.krippendorf@physik.uni-muenchen.de}, {\bf Michael Spannowsky}${}^{\,b,c,}$\footnote{michael.spannowsky@durham.ac.uk},\\
{\small
\vspace*{.5cm}
${}^{a\;}$Arnold Sommerfeld Center for Theoretical Physics, Ludwig-Maximilians-Universit\"at,\\
Theresienstr.~37, 80333 Munich, Germany\\[3mm]
${}^{b\;}$Institute for Particle Physics Phenomenology, Department of Physics, Durham University, South
Road, Durham DH1 3LE, United Kingdom\\[3mm]
${}^{c\;}$ Department of Physics, Durham University, DH1 3LE, United Kingdom

}
\end{center}
\vspace{1cm}
\begin{abstract}

\noindent 
We demonstrate that the dynamics of neural networks trained with gradient descent and the dynamics of scalar fields in a flat, vacuum energy dominated Universe 
are structurally profoundly related. This duality provides the framework for synergies between these systems, to understand and explain neural network dynamics and new ways of simulating and describing early Universe models. 
Working in the continuous-time limit of neural networks, we analytically match the dynamics of the mean background and the dynamics of small perturbations around the mean field, highlighting potential differences in separate limits. We perform empirical tests of this analytic description and quantitatively show the dependence of the effective field theory parameters on hyperparameters of the neural network.
As a result of this duality, the cosmological constant is matched inversely to the learning rate in the gradient descent update.

\end{abstract}

\newpage

\tableofcontents

\section{Introduction}
Effective field theories (EFT) are one of the most powerful paradigms of contemporary physics, which can capture the dynamics in virtually all science areas, thereby covering the vast scales relevant in cosmology, the intermediate scales relevant for living organisms, and the very small scales of particle physics. One aspect of the astounding power of effective field theories is that they do not require a detailed understanding of the microphysical processes at an elementary level to describe the dynamics in question. Instead, effective degrees of freedom and interactions are being utilised to describe physical processes at the relevant scales.

Here, we propose and exemplify how to identify the effective field theory associated with neural network (NN) training dynamics. The network dynamics are characterised by how the neural network output changes during training, i.e.~how the network weights are updated. Deep neural networks (DNN) have entered a revolutionary phase, where their near-term prospect foresees the deployment of sophisticated methods to safety-critical applications, e.g. medical applications, autonomous driving, or financial investments. DNN are often described as black boxes whose features are not 
explainable, i.e.~it is generally not understood what state neural networks 
strive towards after and during training. Because of their intrinsic high degree of non-linearity, the ubiquitous deep learning algorithms, which are at the forefront of the Machine Learning (ML) developments in most areas, still lack transparency and explainability. Thus, being able to improve our understanding of the intrinsic dynamics of neural networks during training is of particular importance for the future development and deployment of DNNs in addressing real challenges. 

In general, a DNN starts from some initial functional representation which changes guided by the loss function. Following the gradient-based optimisation algorithm, aiming to minimise the loss function, it evolves to contain non-linear features which are well-suited for classification or regression tasks. To avoid getting stuck in local minima, gradient descent methods are often augmented. One such method is to modify the weight update by a momentum term \cite{polyak}, an aggregate of gradients, which becomes the exponential moving average of the history of gradients. The parameters of a network are updated regularly, after a discrete number of steps, measured in batches or epochs. When taking the continuous limit for the network's learning-rate equation in the presence of a momentum term, one finds a linear second-order differential equation in the time component. Limiting the network output to a single value, its structure equates a scalar field, defined over the the continuous, potentially multi-dimensional, input 
space, very similar to the equations that govern the dynamics of the early Universe. Thus, it is plausible that such scalar field equations, can describe the training of a sufficiently wide neural network.

It is therefore intriguing to compare the training of a neural network with the evolution of a scalar field in the Friedmann-Lema\^itre-Robertson-Walker (FLRW) background, which is a solution of Einstein's field equations of general relativity. The relatively simple form of these scalar field equations are a direct consequence of the assumption that the Universe is homogeneous and isotropic. However, non-linear features can develop from random fluctuations around a homogeneous time-evolving background (cf.~\cite{Mukhanov:2005sc, Weinberg:2008zzc, Baumann:2009ds} for a concise overview). The time evolution of the scalar background field depends on the gravitational theory underlying the dynamics, which includes in the FLRW case of general relativity whether other energy components dominate the Universe. This is naturally analogous to choices for the optimisation algorithm, most prominently the choice of the learning rate used for the gradient-based update.
 
In this paper, we establish this duality between previously unrelated dynamical systems by demonstrating that the evolution of the neural network follows very comparable dynamics of a scalar field in an FLRW-background, thereby resembling the dynamics in the early Universe. 
Relating both dynamical systems offers the potential for significant synergies drawing from knowledge on the respective process: On the one side, this includes the fact that the formation of structure in our Universe can be modelled very accurately in the EFT framework of $\Lambda$CDM and in similar ways extensions of this standard model in the EFT language are utilised to describe potential additional features which can be tested observationally, a prime example being the plethora of dark matter models which range from ultralight axion-like particles to primordial black holes~\cite{ParticleDataGroup:2020ssz}. On the other hand, the functional dependence of the EFT on hyperparameters of the neural network and the optimiser allow the optimisation of such parameters without having to train multiple networks. This duality also relates scales of different parameters (i.e.~optimisation, training set size, and NN architecture) to one dynamical/analytical framework.

This new EFT perspective, in comparison to previous work on approximate NN dynamics, in particular the neural tangent kernel approach (NTK)~\cite{ntk1,1902.06720,novak2019neural}, allows us to understand and utilise the hierarchies in the dynamics appearing in the NTK formulation. This is to say that we can identify a mode dominating the dynamics, the mean field of the NN, and that the evolution of the other modes is heavily influenced by the evolution of the mean field. Following the EFT framework, we also identify modes which effectively are not evolving and which can be neglected without loosing accuracy in training. This EFT approach allows for a systematic investigation of whether all types of dynamics which can appear in the early Universe have already been constructed in the context of NNs.

 This duality between field theory dynamics and NNs updated with gradient descent also offers a new perspective on describing the observed energy distribution in our Universe. For instance, we find a close relationship between the learning rate and the vacuum energy. Further, as training a neural network with gradient descent is reasonably cheap, it is a compelling question whether our duality between these two systems leads to more efficient simulation algorithms than {\it standard} discretisation of continuous dynamics (e.g.~in lattice simulations). Put differently, it is interesting to see which phenomenological EFT models share dynamics with neural networks. In addition, this NN perspective necessitates the presence of gravitational dynamics in any field theoretical evolution given sufficient excursions in field theory space where gravitational dynamics cannot be decoupled.

To demonstrate a duality between both dynamical systems, we match the evolutionary dynamics in the continuous-time limit in Sec.~\ref{sec:matching}. We perform this matching explicitly for the simple case of a neural network with a single output dimension related to a single scalar field analytically. The scalar field is exposed to a scalar potential associated with the neural network architecture and the loss function. Moreover, the optimisation algorithm is related to the gravitational dynamics of the scalar field. At this stage, we require that the empirical NTK be a valid approximation of the neural network dynamics during training. To make this match with cosmological dynamics, we perform a basis change that diagonalises the empirical NTK. We find that the dynamics of the modes are of the same structure with the difference that the NN modes evolution can be suppressed in the relevant evolution term by the smaller eigenvalues appearing in the NTK, which depends on the choice of the NN architecture and the training data. In Sec.~\ref{sec:matching_quantitatively}, we empirically test this duality by training homogeneous NNs to match the dynamics of a homogeneous Universe and then standard input dependent NNs to match both the dynamics of the homogeneous Universe and the evolution of the perturbation. We utilise standard fully-connected NNs, which are optimised using gradient descent with and without momentum subject to a polynomial loss function resembling of mean squared error and, more importantly, those of standard polynomial inflationary potentials~\cite{Linde:1983gd}. In Sec.~\ref{sec:related_work} we connect our duality with previous directions of work in both ML and physics before concluding in Sec.~\ref{sec:conclusions}.

\section{Analytically matching dynamics}
\label{sec:matching}
Here we match the dynamics of a scalar field in an FLRW background and the continuous-time dynamics of neural networks, which are updated via gradient descent with momentum, closely following the notation in~\cite{1902.06720}.\footnote{Previous work on continuous-time updates include~\cite{Qian, Su}.} To make the connection, we start with the homogeneous and isotropic evolution of a perfect fluid, i.e.~a spatially homogeneous  and isotropic scalar field. The most straightforward realisation is with networks that are homogenous and isotropic with respect to the input. We then relax this homogeneity and istropy assumption and match the dynamics for small perturbations around such a homogeneous and isotropic background.

In our comparison, we must study the NN dynamics on the function space level rather than the `microscopic' NN parameter level. This perspective is in contrast to parameter-based approaches, which share a similar aim to explain the NN dynamics and their learning behaviour~\cite{Roberts:2021fes}.

\subsection{Matching homogeneous and isotropic dynamics}
We are interested in describing the dynamics of a NN $f(t,x,\theta)$ subject to gradient descent
\begin{eqnarray}
f:~\mathbb{R}\times\mathbb{R}^n\times\mathbb{R}^m &\to& \mathbb{R}\\
\nonumber (t,x,\theta)&\mapsto& f(t, x,\theta)~.
\end{eqnarray}
Gradient descent with momentum yields a continuous time differential equation for the update of the neural network which is of second order.
The discrete update equation of weights is
\begin{equation}
\theta_{i+1}=\theta_i+\beta (\theta_i-\theta_{i-1})-\eta~\nabla_\theta{\cal L}({\cal D})|_{\theta=\theta_i}~,
\end{equation}
where $\beta$ denotes the momentum parameter, and $\eta$ the learning rate. ${\cal L}$ is the loss function which we consider to be of the following type:
\begin{eqnarray}
{\cal L}:\mathbb{R}\times\mathbb{R}&\to& \mathbb{R}\\
\nonumber (y,y_{\rm target}) &\mapsto& {\cal L}(y,y_{\rm target})~,
\end{eqnarray}
where we average this expression when it is evaluated over a (training) dataset ${\cal D}$ with the length of the dataset $N=|{\cal D}|.$ Its continuous version is 
\begin{equation}
\ddot{\theta}=\tilde{\beta}\dot{\theta}-\nabla_\theta{\cal L}({\cal D})~,
\end{equation}
where $\tilde{\beta}=(\beta-1)/\sqrt{\eta}$ and the continuous and discrete time are related via $t=i \sqrt{\eta}.$
The update equation for the neural network output is obtained by utilising the chain rule $\dot{f}=\dot{\theta}\nabla_\theta f$, and becomes
\begin{equation}
\ddot f(x)=\tilde{\beta}\dot{f}(x)-\nabla_\theta f(x)\nabla_\theta f({\cal D}) \nabla_f {\cal L}({\cal D})~,
\label{eq:updatenn}
\end{equation}
which can be evaluated locally for each point $x$ of the input space. The (non-local) data used for updating the weights ${\cal D}$ has to be applied when calculating~$\dot{\theta}.$

Before focusing on specific limits in the NN dynamics, let us briefly discuss the scalar dynamics in a homogeneous and isotropic Universe, described by 
\begin{equation}
  \ddot{\phi}+3H\dot{\phi}+V'(\phi)= 0~,~\; \text{where }3H^2=\frac{\dot\phi^2}{2}+V(\phi)~,
  \label{eq:flrw-evolution}
\end{equation}
where $\phi(x,t)=\phi(t)$ denotes the homogeneous and isotropic scalar field and $V(\phi)$ its scalar potential. $H$ is the Hubble parameter which is the ratio of the change of the metric's scale factor and the scale factor itself. Einstein's equations set the Hubble parameter in the case of our scalar field as in \eqref{eq:flrw-evolution}. 

To achieve a matching between the differential equations in \eqref{eq:updatenn} and \eqref{eq:flrw-evolution}, we specialise first to the homogeneous and isotropic case for the input (feature) space, i.e.~$f(x,t)=f(t)$ and then discuss the general case. In this case, the prefactor is defined by $\alpha\equiv\nabla_\theta f(x,t) \nabla_\theta f({\cal D},t).$
In the NTK limit, i.e. the limit where the network is infinitely wide, the prefactor is constrained to its value at the beginning of training, $\alpha(t)=\alpha(t=0)$. Thus, one obtains a constant $\alpha.$ In this limit the matching becomes straightforward as the NN dynamics simplify to
\begin{equation}
\ddot{f}-\tilde{\beta}\dot{f}+V'_{\rm eff}(f)= 0~,
 \label{eq:nnhomogeneousapproximated}
\end{equation}
where the potential $V_{\rm eff}$ is specified in due course.

The second approximation which we make -- this time on the FLRW side of the duality -- is that we assume $H$ to be dominated by the vacuum energy, a parameter which is eliminated from $V'(\phi).$ Given these approximations, an effective potential of the following form matches the dynamics of scalar fields in a homogeneous and isotropic Universe with those of a neural network
\begin{equation}
V_{\rm eff}=\alpha {\cal L}+V_0~,~{\rm where}~V_0=\frac{\tilde{\beta}^2}{3}~,
\end{equation}
where the vacuum energy $V_0$ is added to ensure a matching of the two respective friction terms. Thus, we find that the network dynamics resemble the dynamics of a vacuum energy dominated Universe, i.e.~$3H^2\approx V_0.$

It is rather remarkable that the vacuum energy has the interpretation of the combination of momentum parameter and learning rate in the neural network dynamics. To our knowledge, this connection is new. It is also remarkable that in this picture, it is clear that a change in the neural network architecture can change $\alpha$, which in turn determines the training dynamics, e.g.~whether the training dynamics are vacuum energy dominated or not, and this direct matching applies.

We also note that a scalar field in our Universe can change the optimiser parameters dynamically. A systematic investigation of the exact matching between optimiser algorithms and standard phases of cosmological evolution, e.g.~domination of the kinetic terms, is left for future work. We note that a parametrically small cosmological constant in this dual picture corresponds to a parametrically large learning rate in the optimisation algorithm. This matching of optimiser parameters and the vacuum energy matches respective 'global', i.e.~non-local, quantities.

Finally, let us comment on the scales which have been set in this duality. In particular, the Hubble scale is relevant for the time evolution and the appropriate scale in the spacetime metric. The duality relates to the previously unrelated scales associated with the input data and the time steps. In particular, the Hubble time is $1/H$, and it corresponds to the only time-scale in the neural network. This is in contrast to our Universe where additional dynamics are taking place beyond the ones related to {\it training}. 

\subsection{Matching perturbations}

In early Universe physics, quantum fluctuations around the classical background can grow large and are responsible, in the standard cosmological picture, for structure formation. The equations for small perturbations $\delta\phi\ll 1$ can be obtained by considering the entire equations of motion with the ansatz $\phi+\delta\phi.$
In NNs, inhomogeneities are essential for successfully addressing tasks, e.g.~the classification of images. To establish the similarities between both dynamics, we focus on neural networks which have small perturbations where we specify the appropriate size in due course. 

\subsubsection*{NN perturbations}

We start with the continuous time version of the dynamics 
\begin{equation}
 \ddot{f}(t,x)-\tilde{\beta}\dot{f}(t,x)+\Theta(t,x,X)\nabla_{f(X)}{\cal L}=0~, \label{eq:ntkcontinuous}
\end{equation}
where $\Theta(t,x,X)$ denotes the empirical neural tangent kernel at a given time $t$ during the learning process.\footnote{The linearisation in the NTK limit takes this to be evaluated at the beginning of training.} In the classical limit we focused on a constant spatially homogeneous and isotropic value for the empirical neural tangent kernel $\Theta(X,X)\approx \alpha.$ Now, we are interested in taking into account the deviation from this approximation $f(t,x)=\bar{f}(t)+\delta f(t,x).$

To derive the equations, let us write the last term in~\eqref{eq:ntkcontinuous} more explicitly:
\begin{eqnarray}
 \Theta(t,x,X)\nabla_{f(X)}{\cal L}&=&\sum_{y\in X} \Theta(t,x,y)\nabla_{f(y)}{\cal L} \nonumber \\
 &=&\sum_{y\in X} \Theta(t,x,y){\cal L}'(f(y)) \nonumber \\
 &=&\sum_{y\in X} \Theta(t,x,y)\left[{\cal L}'(\bar{f})+{\cal L}''(\bar{f})~\delta f(y)\right]~, \label{eq:pert}
\end{eqnarray}
where we have assumed in the first line that our loss function can be expressed as a sum over individual data points, i.e. point-wise local contributions over the input space. The second line is a simple rewriting and the third line is a linear expansion in $\delta f$, which is exact for a quadratic loss function.
In general, the approximation in~\eqref{eq:pert} remains valid for small perturbations $\delta f$ where the deviation from the exact NTK result provides a first criterion on when perturbations can be considered small.

To make contact with the previous mean-field evolution, we perform a basis transformation to diagonalise the neural tangent kernel, decoupling the modes' evolution associated with the respective eigenvalues. In particular we find in our experiments the largest eigenvalue corresponds to the mean field (i.e.~the uniform evolution) $\alpha$ from before.
\begin{eqnarray}
 \Theta(t,x,X)\nabla_{f(X)}{\cal L}&=&\Theta_{ij} {\cal L}'_j \nonumber \\
 &\to& A_{ik} \Theta_{kl} (B^T)_{lj} A_{jm}{\cal L}'_m
\end{eqnarray}
where $A$ and $B$ are matrices which can be used to diagonalise $\Theta(t=0,X,X),$ $B^T A={\bf 1},$ and $\Theta(x_i,x_j)=\Theta_{ij}.$ We construct $A$ and $B$ from the normalised eigenvectors $v_i$ of $\Theta_{ij}$ as
\begin{eqnarray}
 && A=\left(\begin{array}{c}
          v_1^T\\
          v_2^T\\
          ...\\
          v_N^T
         \end{array}
 \right)~,~ B=\left(\begin{array}{c}
           v_1^T\\
          v_2^T\\
          ...\\
          v_N^T
         \end{array}
 \right),~v_i^T\cdot v_j=\delta_{ij}~,\\
 && A~\Theta~B^T={\rm diag}\left(\lambda_1,\ldots,\lambda_N\right).
\end{eqnarray}
Using this basis transformation we can now define what is meant with perturbations around the mean field, namely we utilise:
\begin{equation}
f_i = \sqrt{N}v_1\bar{f}+\delta f_i~, \text{ and }{\cal L}_i=\frac{m^2}{N}(f_i-f_0)~,
\end{equation}
where we used a quadratic loss function with the minimum at $f_0$ for illustrative purposes.

This basis change and this definition of perturbations not only decouples the dynamics of modes but also singles out the mean field evolution. 
Introducing $\delta \tilde{f}_i=v_i^T\cdot \delta{\bf f}$, this gives the following equations for the mean and the perturbation evolution
\begin{eqnarray}
 0&=&\ddot{\bar{f}}+\tilde{\beta}~\dot{\bar{f}}+\frac{\lambda_1}{N}~{\cal L}'(\bar{f})~,  \label{eq:meanevolution}\\
 0&=&\delta\ddot{\tilde{f}}_i+\tilde{\beta}~\delta\dot{\tilde{f}}_i+\frac{\lambda_i}{N}~\delta\tilde{f}_i~{\cal L}''(\bar{f})~.
 \label{eq:perturbationevolution}
\end{eqnarray}
More details on the derivation can be found in Appendix~\ref{app:derivation}.

\subsubsection*{Scalar field perturbations}
Let us now compare the above equations with the general equations of motion for a scalar field in a FLRW background which read
\begin{equation}
\label{eq:start}
 \ddot{\phi}+\nabla^2\phi+3H\dot{\phi}+V'(\phi)= 0~.
\end{equation}
In the mentioned scenario, where the scalar field can be expanded into a universal background and a small perturbation $\phi(t,x)=\phi(t)+\varphi(t,x)$ equation \eqref{eq:start} can be separated into 
\begin{eqnarray}
 \ddot{\phi}+3H\dot{\phi}+V'(\phi)&=& 0~, \label{eq:mean} \\
 \ddot{\varphi}+\nabla^2\varphi+3H\dot{\varphi}+V''(\phi)\varphi&=&0~, \label{eq:pert_cosmo}
\end{eqnarray}
where in the last equation we are working to linear order in the perturbations ($V'(\phi+\varphi)\approx V'(\phi)+\varphi V''(\phi)$).  Equation \eqref{eq:mean} corresponds to the homogeneous and isotropic background field equation whose dynamics we have already matched. 
Instead, \eqref{eq:pert_cosmo} is identical to \eqref{eq:perturbationevolution} when neglecting the spatial gradient term which is a valid approximation of long wavelength/small momentum perturbations and assuming $\lambda_i=\lambda_1.$

\section{Quantitatively matching dynamics}
\label{sec:matching_quantitatively}
We now turn to experimentally test our analytic framework from the previous section by confronting the actual neural network dynamics with our predicted dynamics. In this context, the aim of our experiments is two-fold:
\begin{enumerate}
 \item We would like to assess how accurate our approximations for the mean field and small perturbations around it are, i.e.~when are Equations~\eqref{eq:nnhomogeneousapproximated},~\eqref{eq:meanevolution}, and~\eqref{eq:perturbationevolution} capturing the NN dynamics.
 \item We are interested in quantitatively determining the dependence of the effective field theory potential on the neural network parameters, i.e.~determining $\alpha$ and $\tilde{\alpha}$ empirically.
\end{enumerate}
\begin{figure}[t]
\begin{center}
 \includegraphics[width=0.7\textwidth]{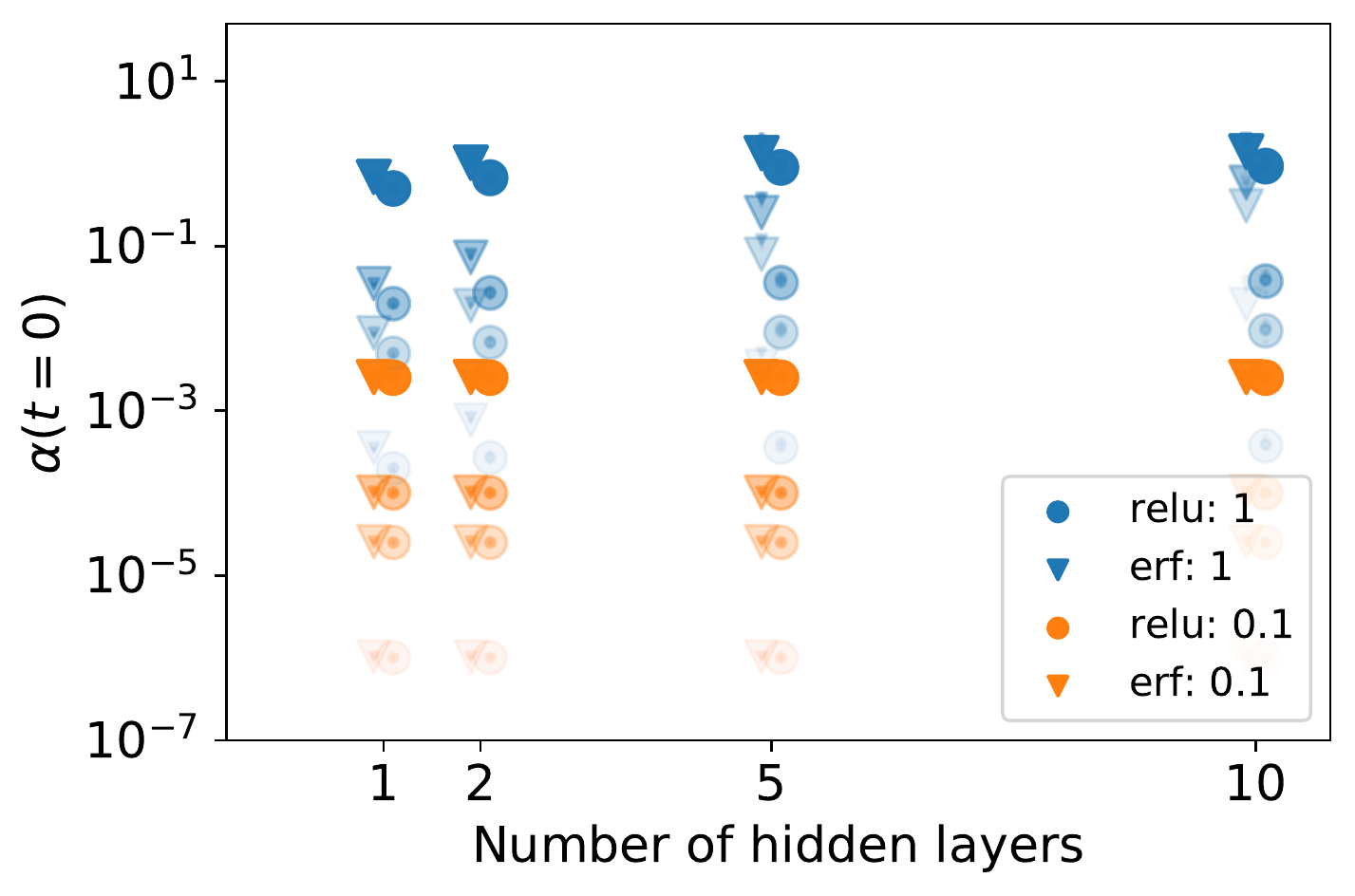}
 \end{center}
 \vspace{-0.5cm}
 \caption{The values of $\alpha$ at initialisation in our hyperparameter scan. The weight initialisations are $1$ (blue) and $0.1$ (orange). We have varied the width of the hidden dense layers among the values $\{100,1000,10000\}$ which is shown with increasing sizes of the marker. Different bias initialisations are taken from $\{0.001,0.005,0.01,0.05,0.5\}$ with increased visibility for larger initialisation. As indicated, we utilised {\it erf} and {\it relu} activation functions and the number of dense layers.} \label{fig:alphadependencenn}
\end{figure}
Our experiments are greatly facilitated by the empirical implementation of the NTK~\cite{novak2019neural}.\\
 For our experiments we choose a simple polynomial loss function which features a minimum away from the initial values of the NN,
\begin{equation}
{\cal L}(f)=\frac{m^2}{2}\left(f-f_0\right)^n~,
\label{eq:loss}
\end{equation}
where $n$ is even and we focus on the cases $n=\{2,4\}$, vary $f_0=\{1,10\}$ and scale the potential as $m=\{0.01, 0.1, 0.2\}$. Besides the simplicity on the neural network side, this is to match with standard field theory potentials, where $n=2$ corresponds to the case of chaotic inflation~\cite{Linde:1983gd}. For our NN, we use fully connected neural networks with different widths of the hidden units, several hidden units, activation functions ({\it relu} and {\it erf}), and the standard deviation of the normal distributions used for the weight initialisations of our dense layers.
To study homogeneous and isotropic neural networks, i.e.~$f(x,t)=f(t),$ we simply evaluate our networks at a single point. In this case the empirical neural tangent kernel is $\Theta(t,x,y)=\alpha.$

Figure~\ref{fig:alphadependencenn} shows the dependence of $\alpha$ as a function of network parameters. The largest effect results from changing the initialisation distributions, followed by smaller differences for the activation function and the number of hidden layers. We find that {\it erf} for activation function leads to larger values than {\it relu}. In contrast, the width has only a smaller effect.

\begin{figure}[t]
\begin{center}
 \includegraphics[width=0.7\textwidth]{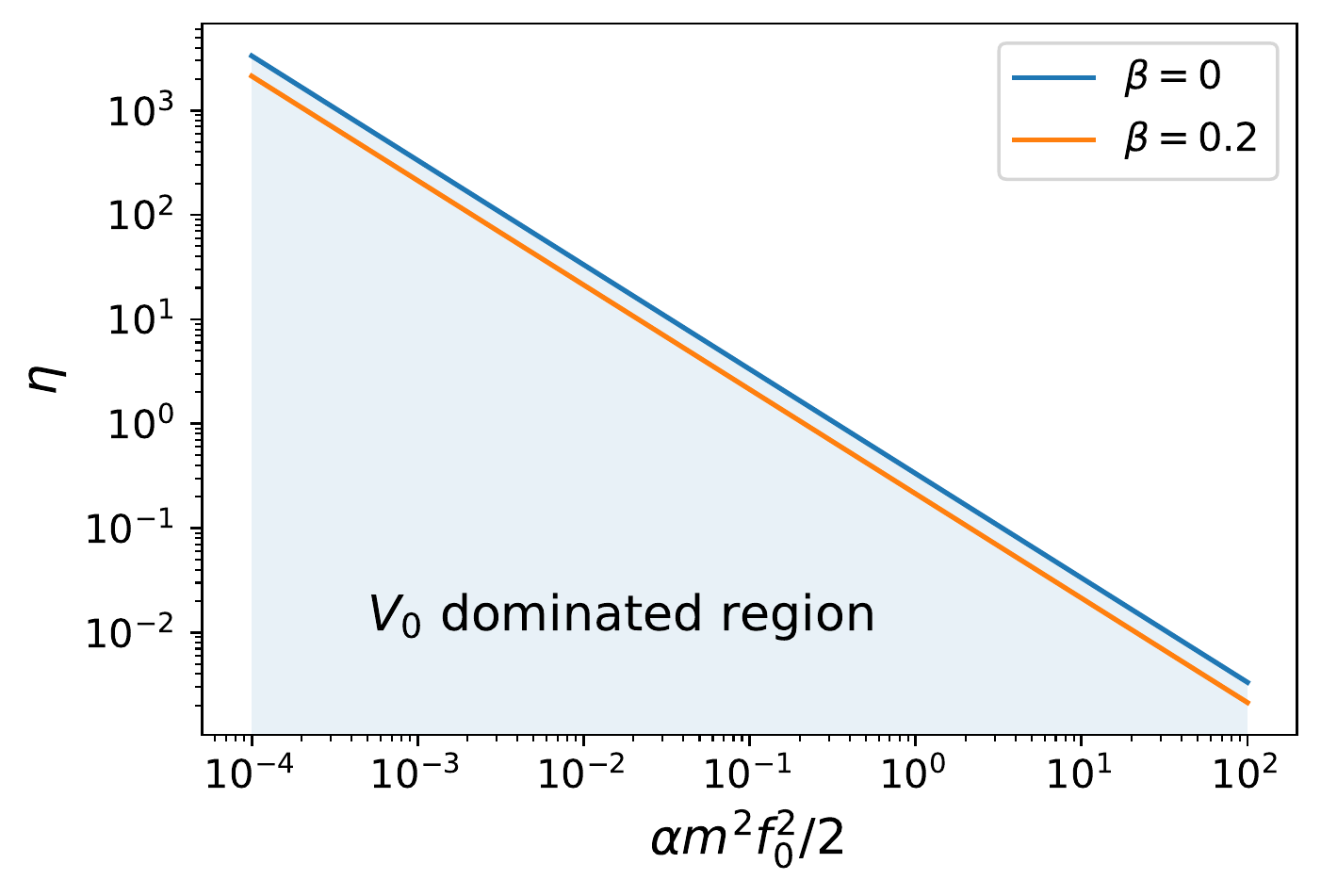}
 \end{center}
  \vspace{-0.5cm}
 \caption{Region of $\eta$ where $V_0$ is larger than the contribution from the other potential terms. When including momentum the allowed region shrinks. The line is set by $V_0=\alpha{\cal L}.$}\label{fig:lrallowed}
\end{figure}

To match the dynamics throughout the entire training, we need to satisfy $\alpha {\cal L}(f)\ll V_0.$ Given a choice for the NN and the loss function parameters $m$ and $f_0,$ this condition implies a maximally allowed learning rate which is reduced when adding non-vanishing momentum. Figure~\ref{fig:lrallowed} shows this dependence and indicates that for an extensive range of parameters, generic values of the learning rates are allowed. This condition relates the learning rate with the scale of the effective potential. In turn, this implies, for instance, that an architecture choice with a larger $\alpha$ requires a lower learning rate to keep this approximation of vacuum energy domination valid.

\begin{figure}[t]
\begin{center}
\includegraphics[width=0.7\textwidth]{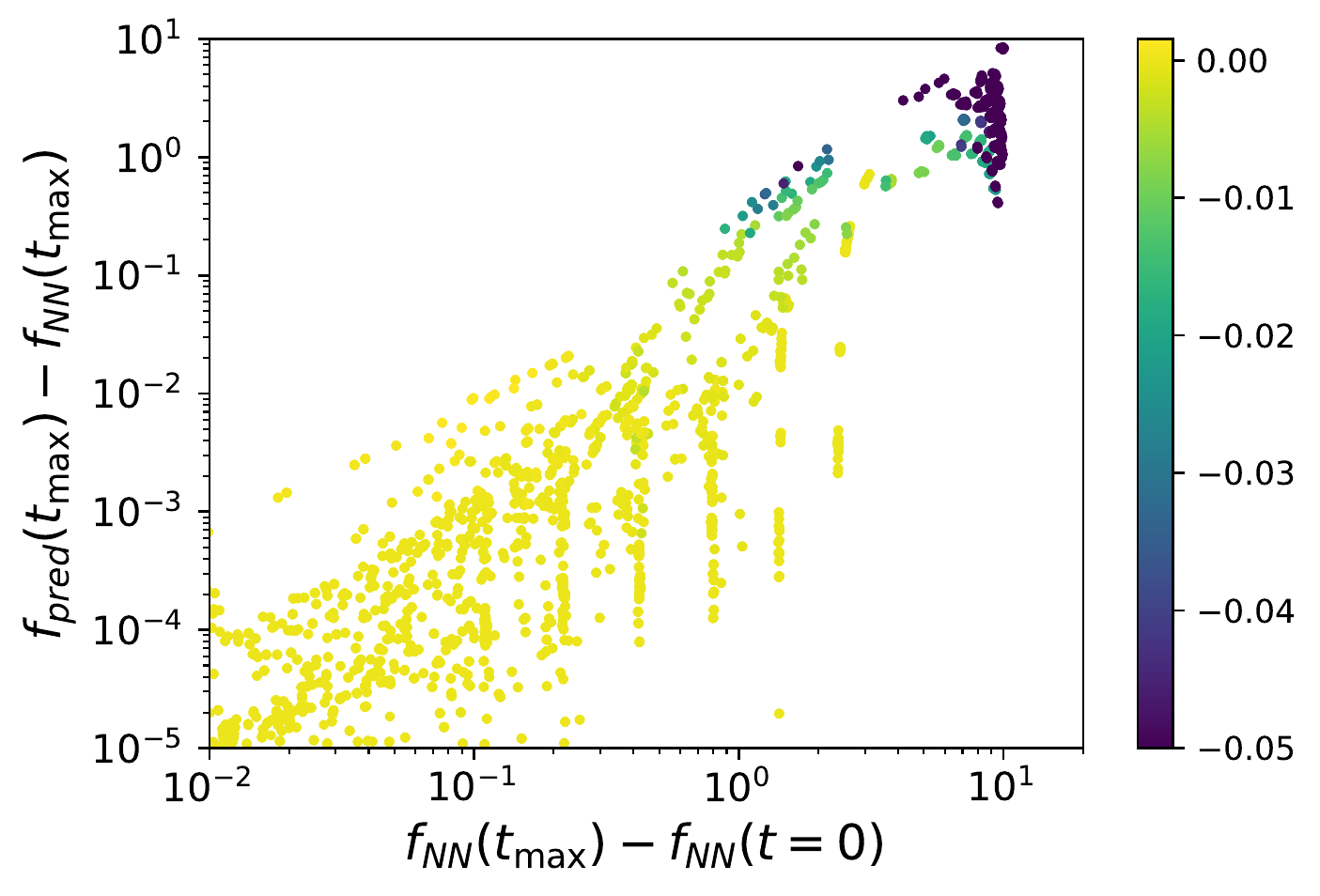}
\end{center}
 \vspace{-0.5cm}
\caption{Comparison of the NN evolution and the predicted evolution according to Eq.~\eqref{eq:meanevolution}. The observed discrepancy is related to the actual evolution in function space during training and found to be significantly smaller. The colour corresponds to the difference of $\alpha$ at the beginning and the end of training $\alpha(t_{\rm max})-\alpha(t=0).$
\label{fig:uniformcomparison}}
\end{figure}

Having chosen hyperparameters, we train our networks explicitly and compare them with the predicted dynamics. Figure~\ref{fig:uniformcomparison} shows the deviation of the predicted dynamics to the actual dynamics at the end of training (i.e.~after $1000$ epochs) in comparison to the difference of the NN output at the beginning and the end of training\footnote{We show results for the following hyperparameter choices: $\{1,2,5\}$ hidden layers with widths $\{100,1000\},$ $\{\text{\it erf,relu}\}$ activations, and initialisations $\{(1,0.05),(0.1,0.05)\}$ learning and momentum rate combinations $\{(0.1111,0.00),(0.4444,0.00),(0.0011,0.5),(0.0278,0.5),(0.0278,0.5),(0.1111,0.5)\}$. The loss function choices were as mentioned before.}. We find excellent agreement across our hyperparameter search, and the most significant discrepancy is observed for large differences in $\alpha,$ which corresponds to the colour coding of the individual points ($\alpha(t_{\rm max})-\alpha(t=0)$). 

\begin{figure}[t]
\begin{center}
\includegraphics[width=0.7\textwidth]{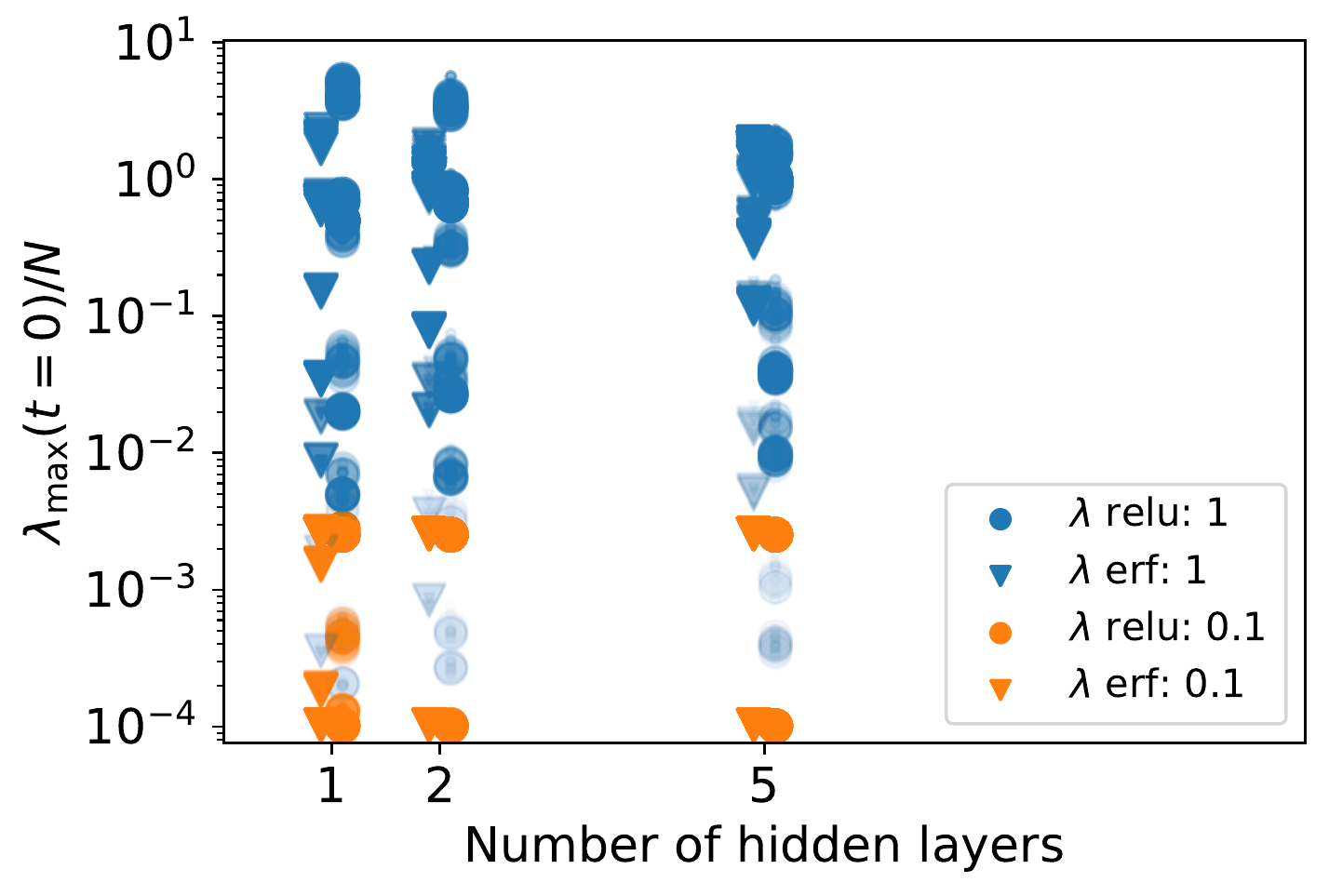}
\end{center}
 \vspace{-0.5cm}
\caption{The values of the largest eigenvalues of the NTK-kernel divided by the number of points in the interval in our hyperparameter scan. The marker coding corresponds to the same coding as in Figure~\ref{fig:alphadependencenn}.\label{fig:lambda_distribution}}
\end{figure}

\begin{figure}
\begin{center}
\includegraphics[width=0.48\textwidth]{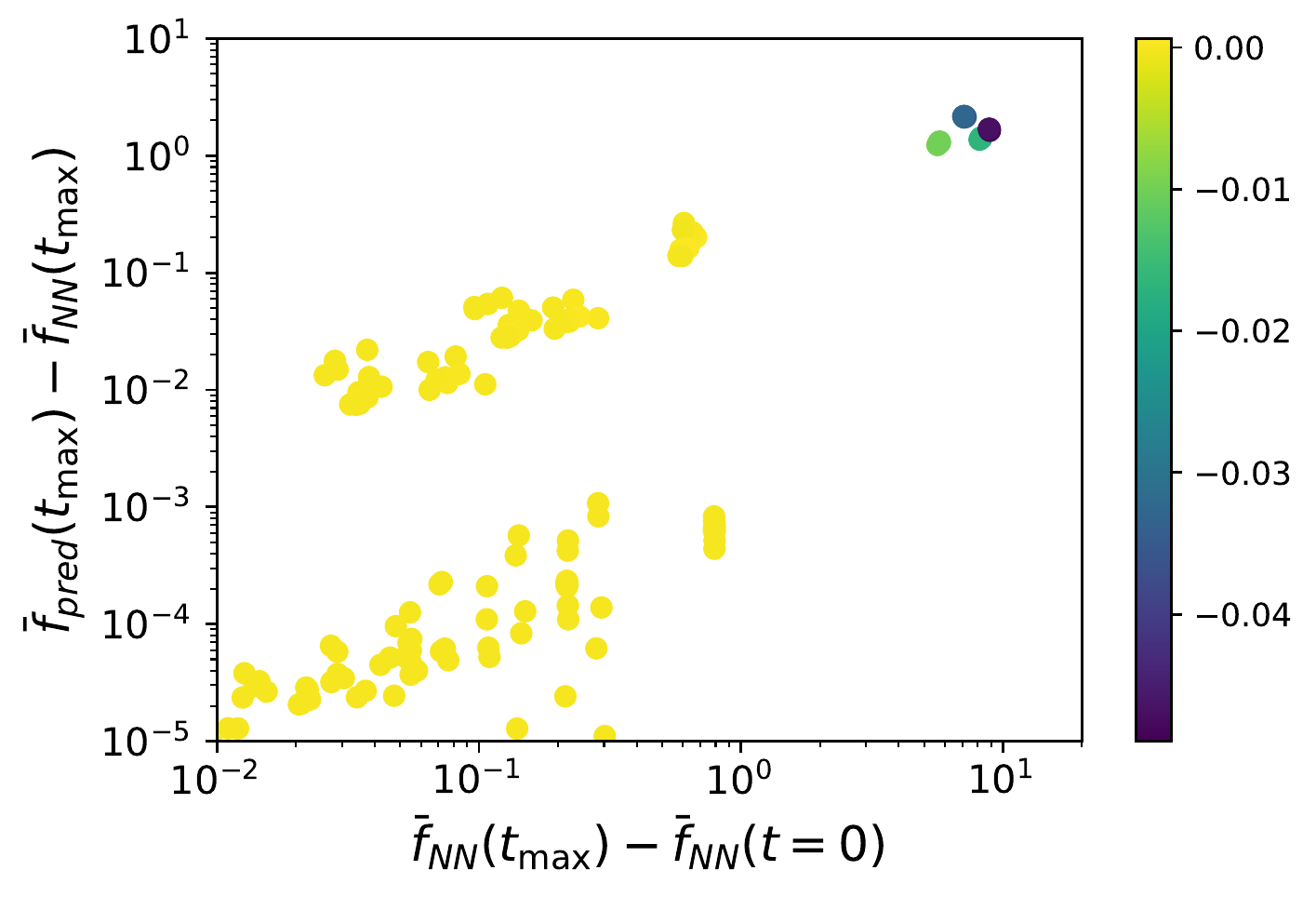}
\includegraphics[width=0.48\textwidth]{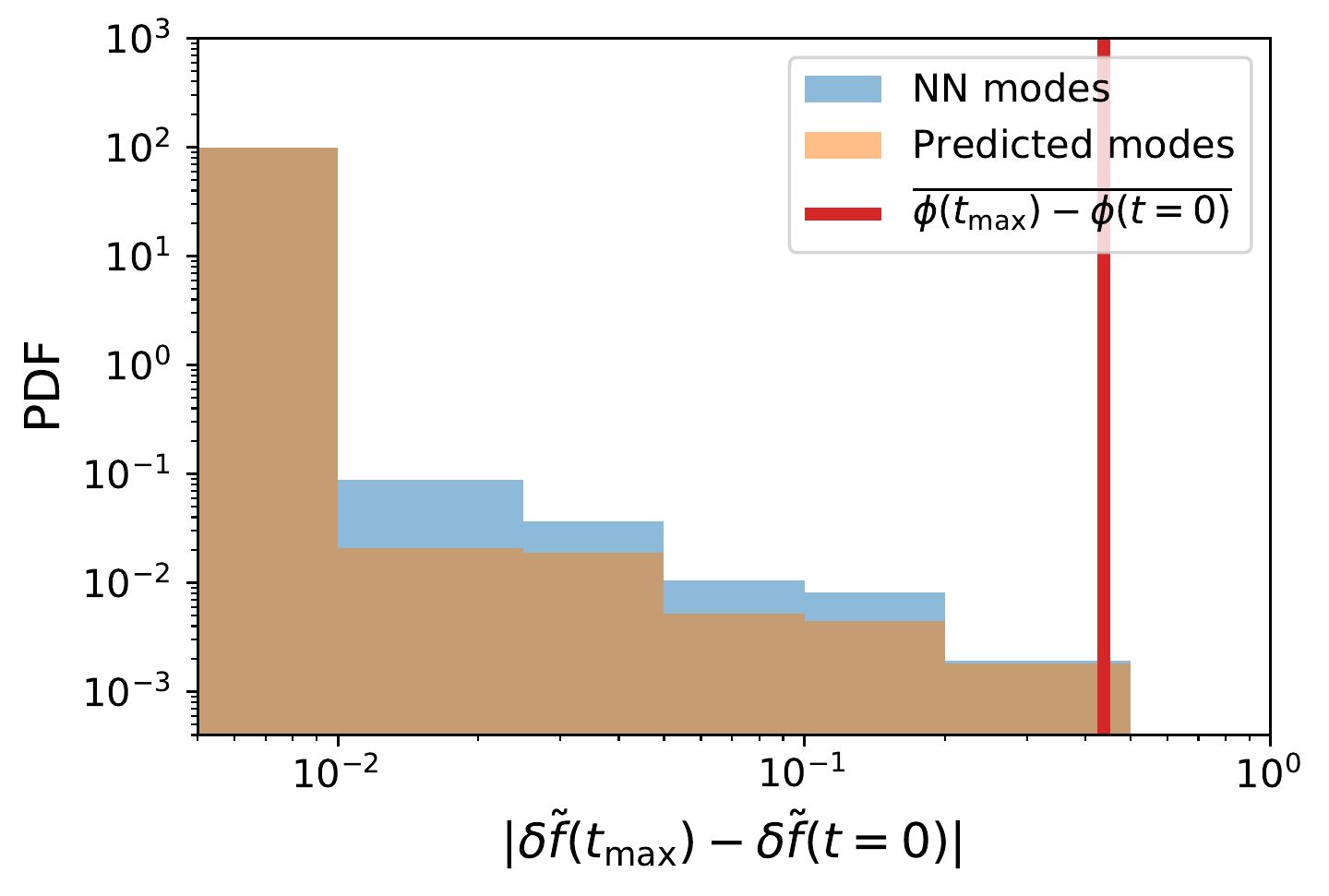}
\end{center}
 \vspace{-0.5cm}
\caption{{\bf Left:} The difference between the predicted and observed mean evolution using Eq.~\eqref{eq:meanevolution} in comparison to the actual mean evolution. {\bf Right:} Histogram of the modes at the beginning and the end of the training interval for the actual NN modes and the predicted modes. The read line shows the hierarchically larger mean value of the respective mean evolutions. \label{fig:inhomogeneousevolution}}
\end{figure}

For networks with input dependence, we choose a one-dimensional input and data from a finite size interval around zero for simplicity, particularly because it allows for visualisation. The interval is taken as $0.1,1$ times the respective Hubble length, set by $3\sqrt{\eta}/(1-\beta)$ and we use $100$ equally spaced points from this interval.
At initialisation, the network varies across the training interval, where the variation depends on the hyperparameter choices and the input interval. The distribution of the largest eigenvalue is shown in Figure~\ref{fig:lambda_distribution} which is very similar to the distribution of $\alpha$ observed in the uniform networks.
We note a large hierarchy between the first and the following eigenvalues (in particular the higher eigenvalues), which is in agreement with previous observations in~\cite{ntk1}. It is this suppression that leads to a freezing of the perturbations as predicted subsequently from our evolution of the perturbations in \eqref{eq:perturbationevolution}. In contrast, the mean evolution, corresponding to the largest eigenvalue, is observed quantitatively as the previous homogeneous and isotropic evolution. The most significant discrepancies are related to hyperparameters with the most considerable deviations in $\alpha$ during training. Figure~\ref{fig:inhomogeneousevolution} compares the predicted mean evolution with the actual mean evolution on the left side.\footnote{This scan involved NNs with one and two hidden layers, $\{\text{\it relu,erf}\}$ activations, $1000$ hidden units, and activations $\{(1,0.05),(0.1,0.01)\}$. We utilised a learning rate of 0.1111 and no momentum.} On the right side, we show the distribution of the observed discrepancies at the end of training between the observed and the predicted mode evolution. We observe the hierarchical separation between the evolution of the perturbation and the mean evolution.
All networks have been trained for $1000$ epochs.

\section{Related work}
\label{sec:related_work}
Various related approaches have been pursued on the physics and the NN side of the duality in recent years.

Undeniably our work builds upon the description of neural network dynamics proposed in the context of the NTK, which was proposed in~\cite{ntk1} and refined in~\cite{1902.06720}. Explicit realisations of the empirical NTK which we utilise can be found here~\cite{novak2019neural}. Our work connects the NTK framework into the physical framework of effective field theories in the early Universe, i.e.~scalar fields in an expanding Universe governed by general relativity.

The neural network output averaged over different samples from the underlying weight distribution can be studied. In appropriate limits, these networks become Gaussian processes~\cite{neal2012bayesian, lee2017deep, matthews2018gaussian, novak2018bayesian, garriga2018deep}. Such Gaussian processes are closely related to the situation when one wants to understand the ensemble of quantum fluctuations of scalar field theories as discussed in~\cite{Halverson:2020trp, Maiti:2021fpy, Erbin:2021kqf,Halverson:2021aot}. Related to our work, we utilise the same interpretation of a neural network as a scalar field. Their approach paves the way for universal statements about a particular architecture, i.e.~it can characterise the size of perturbations. In~\cite{Luo:2021nab}, the loss function and gradient descent updates are included following the NTK procedure. In our work, we link the loss function and the gradient descent update to physical processes in the dual description for the first time. In particular, we propose the distinction between mean-field and perturbation evolution. 

 In contrast to our function-space perspective, \cite{Roberts:2021fes} develops a framework to understand the neural network dynamics by focusing on the dynamics of the individual neural network parameters. This allows to identify scaling such as with the depth and the width of neural networks, i.e.~in our language a microscopic analytic framework to calculate coefficients such as $\alpha.$ Such calculation of effective field theory coefficients from microscopic descriptions is a standard procedure in effective field theory. Here, we opt for the "cheap" option of determining the coefficients experimentally. Along similar lines~\cite{Dyer:2019uzd} presents a framework for understanding neural network dynamics from a parameter space perspective. 
 
Feature learning on the ML side can also be studied using the tensor approach put forward in~\cite{yang2021feature}, following earlier work in this direction~\cite{yang2021tensor,yang2020scaling,yang2020tensor,yang2021tensor3}. Finally, on the cosmology side, feature learning corresponds to understanding the dynamics of the perturbations (cf.~\cite{Mukhanov:1990me} for an early overview).

On the optimiser side, various other algorithms improve performance (e.g.~using Adam optimiser) where the changed update equation changes the differential equation. However, a detailed comparison of different optimisation algorithms, notably including recent physics-inspired ones \cite{DeLuca:2022brp} is beyond the scope of our paper.

Emergent field theory dynamics, including the dynamics of gravity, have been studied from various perspectives in the past. This includes the work of analogue gravity (cf.~\cite{Barcelo:2005fc} for a review), the emergent cosmological dynamics in group field theory~\cite{Gielen:2013kla}, and in holographic setups~\cite{Maldacena:1997re,Verlinde:2010hp}. Our work adds a new dual system to gravitational dynamics. Similarly to other ``dual approaches,'' it is fascinating to analyse and numerically understand gravitational systems' features.

Although the presence of dark energy in our Universe is by now well-established, its theoretical explanations via field theories often suffer from the hierarchically small positive vacuum energy required for consistency with observation in comparison to other natural energy scales of fundamental physics (cf.~\cite{Burgess:2013ara} for an overview of the problem and current approaches). Here the scale of the vacuum energy is related to the parameters of the optimiser. In particular, it is inversely related to the learning rate, which opens a new direction for model building.
 
Extensive cosmological simulations evolve a large number of particles subject to gravitational interactions (e.g.~\cite{gadget}), which build the basis for simulating the currently observed matter distribution in our Universe. Cosmological models at earlier times, in particular dynamics after inflation or during periods of matter domination, often require lattice simulations, utilising software such as LatticeEasy~\cite{Felder:2000hq}. However, both types of simulations are resource-intensive, and this limits the capability of comparing differences in structure formation due to different microscopic models of gravity and particle physics (e.g. other dark matter models). 

\section{Conclusions and outlook}
\label{sec:conclusions}
It seems rather remarkable that the dynamics of neural networks trained with gradient descent are tightly connected with a field theory of gravitational dynamics. It is the interplay of all components in the ML setup, the training data, the NN architecture, the loss, and the optimiser which are required for this duality. As all of these components are naturally required in the NN learning setup, it seems intriguing that the physical system matching such dynamics requires gravitational dynamics, and a purely non-gravitational system (i.e. where gravitational dynamics decouple) does not capture the entire space of dynamics. We find that it is the microphysical model of the optimiser which sets interesting gravitational parameters, particularly the vacuum energy.

This raises several intriguing questions for future investigations:
\begin{itemize}
\item Which early Universe models can be described by NNs? How can we resemble the existing classes by adapting the optimiser, and which other dynamics -- not yet studied in the context of early Universe model building -- can be obtained? On the physics side, this raises another question: how unique are the gravitational dynamics of general relativity, i.e.~to which extent can this duality hold in modified theories of gravity. Finally, on the ML side, to which empirical modifications of the NN training process do such changes in the physics model lead?
\item On the ML side, the physics perspective of the duality allows us to design novel physics-inspired optimisers in the future. They will be inspired by the ``optimisation'' observed in physical systems.
\item How can we extend the current duality beyond the case of a Universe with vacuum energy domination? At this stage, our duality is shown for a small subset of cosmological dynamics, i.e.~the case of vacuum energy domination. It would be exciting to extend this duality in the first step to the standard instances of matter and radiation dominated Universes. However, this requires a modification of the optimisers used for the gradient descent update, which we leave for future work.
\item Neural network dynamics show behaviour similar to that of phase transitions, e.g.~completely different features at initialisation and the end of training, or, put in the language of effective field theories different collective modes dominate the estimation of neural networks. Such phase transitions have happened throughout the evolution in the early Universe wherein different phases of other effective field theories can be used to capture the physics. To trace the cosmological evolution throughout such phase transitions is intrinsically tricky, and lattice simulations can become expensive when high resolutions are required. It will be extremely interesting to trace such phase transitions through our duality. The neural network perspective offers a flexible approach that allows capturing both features through the capabilities of the neural network.
 \item The spectrum of the NTK is dependent on the initialisation and the choice of neural network architecture. It is intriguing to investigate how to match architectures to the scope of perturbations used in cosmology. This offers a new perspective on the initial conditions in the early Universe, i.e.~how are features after training dominated by the choice of initial conditions. In addition it is very interesting to identify situations where more than one eigenvalue remains large, corresponding to more than one background field involved in the dynamics.
 \item This duality provides analytic control on why different architectures can lead to different perturbation growth and how feature learning depends on hyperparameter choices. This is essentially captured by the second derivative of the loss function which sees the mean field act like an external force on a harmonic oscillator (e.g.~allowing for parametric resonance).
\end{itemize}

We hope to return to some of these questions in the not so distant future, shedding more light on the tantalising question: {\it Is our Universe a neural network trained with gradient descent?}

\appendix

\section{Derivation of perturbation equation}
\label{app:derivation}
Here we provide some helpful identities and a detailed derivation of Eqs.~\eqref{eq:meanevolution} and~\eqref{eq:perturbationevolution}. Using the decomposition of our NN into mean field and perturbations the derivatives of the loss function can be decomposed as follows:
\begin{eqnarray}
 f_i &=& \sqrt{N} v_1 \bar{f}+\delta f_i~,\label{eq:perturbationdefinition}\\
 \nonumber {\cal L}_i&=&\frac{m^2}{N}(f_i-f_0)=\frac{m^2}{N}(\bar{f}-f_0)+\frac{m^2}{N}\delta f_i=\frac{1}{N}{\cal L}'(\bar{f})+\frac{1}{N}{\cal L}''(\bar{f})~\delta f_i\\
 &=&\frac{(v_1)_i}{\sqrt{N}}{\cal L}'(\bar{f})+\frac{1}{N}{\cal L}''(\bar{f})~\delta f_i~,\\
 {\cal L}''_i&=&\frac{m^2}{N}=\frac{1}{N}{\cal L}''(\bar{f})~,
 \end{eqnarray}
 Applying the basis transformations on the individual terms leads to
 \begin{eqnarray}
 A_{kl}\ddot{f}_l&=&\left(\begin{array}{c}
                           \sqrt{N}\ddot{\bar{f}}+\left(v_1^T\cdot \delta \ddot{\bf f}\right)\\
                           \left(v_\mu^T\cdot \delta \ddot{\bf f}\right)
                          \end{array}\right),\label{eq:approx2}\\
A_{kl}{\cal L}'_l&=&\left(\begin{array}{c}
                           \frac{1}{\sqrt{N}}{\cal L}'(\bar{f})+\frac{1}{N}\left(v_1^T\cdot \delta {\bf f}\right){\cal L}''(\bar{f})\\
                           \frac{1}{N}\left(v_\mu^T\cdot \delta {\bf f}\right){\cal L}''(\bar{f})
                          \end{array}\right).\label{eq:approx}
\end{eqnarray}
Combining these results, we have
\begin{eqnarray}
 0&=&A_{kl}\ddot{f}_l+\tilde{\beta}A_{kl}\dot{f}_l+A_{kl}\Theta_{lm}B^T_{mn}A_{no}{\cal L}'_o\\
 \nonumber &=&\left(\begin{array}{c}
                           \sqrt{N}\ddot{\bar{f}}+\left(v_1^T\cdot \delta \ddot{\bf f}\right)\\
                           \left(v_\mu^T\cdot \delta \ddot{\bf f}\right)
                          \end{array}\right)+\tilde{\beta}\left(\begin{array}{c}
                           \sqrt{N}\dot{\bar{f}}+\left(v_1^T\cdot \delta \dot{\bf f}\right)\\
                           \left(v_\mu^T\cdot \delta \dot{\bf f}\right)
                          \end{array}\right)\\ &&+\left(\begin{array}{c}
                           \frac{\lambda_1}{\sqrt{N}}{\cal L}'(\bar{f})+\frac{\lambda_1}{N}\left(v_1^T\cdot \delta {\bf f}\right){\cal L}''(\bar{f})\\
                           \frac{\lambda_\mu}{N}\left(v_\mu^T\cdot \delta {\bf f}\right){\cal L}''(\bar{f})
                          \end{array}\right).
\end{eqnarray}
This leads directly to the Eqs.~\eqref{eq:meanevolution} and~\eqref{eq:perturbationevolution}.
\section{Mean field predictions for inhomogeneous networks}
\begin{figure}
\begin{center}
\includegraphics[width=0.8\textwidth]{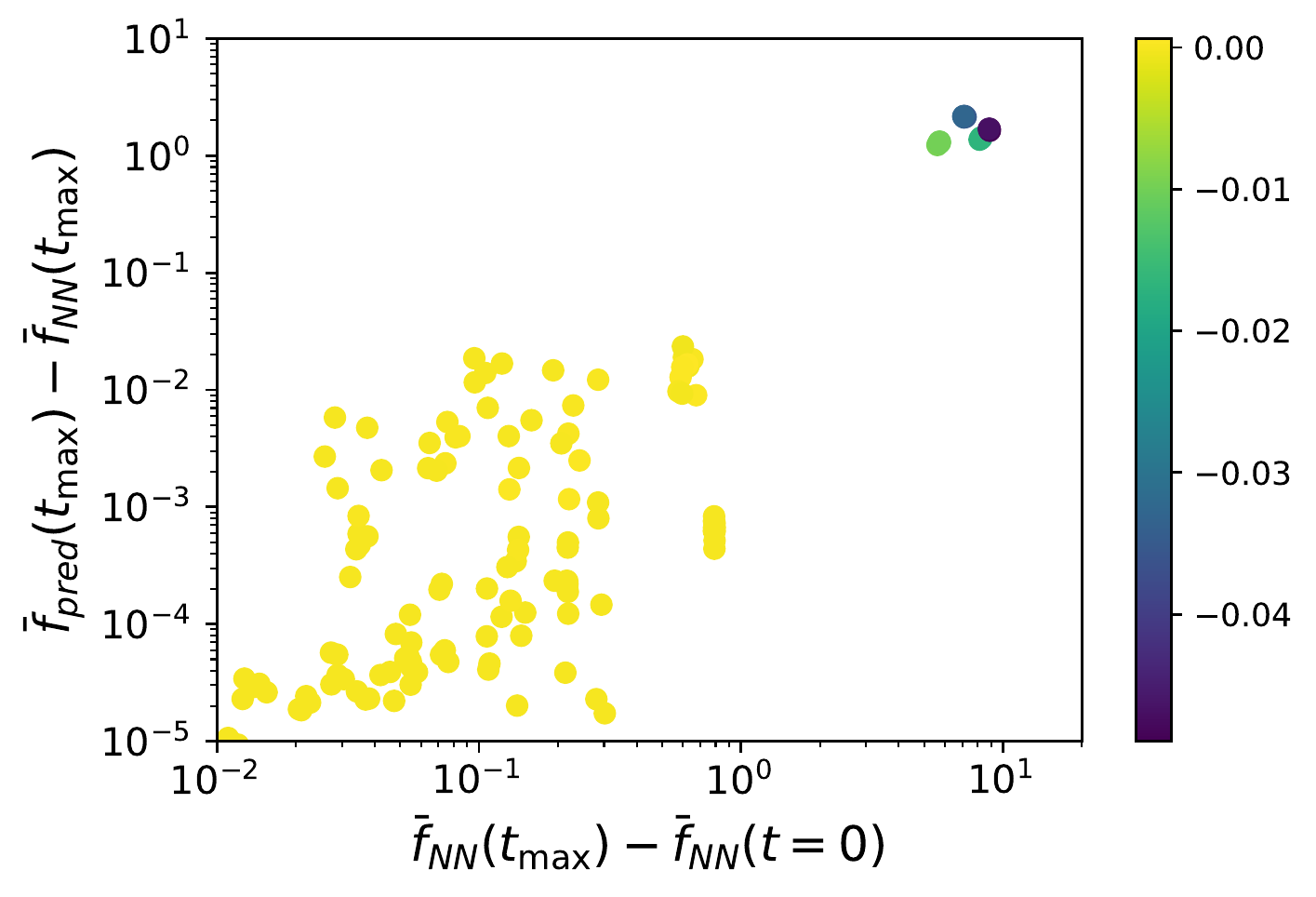}
\end{center}
\caption{The difference between the predicted and observed mean evolution using Eq.~\eqref{eq:nnhomogeneousapproximated} as the evolution for the inhomogeneous NNs used for Figure~\ref{fig:inhomogeneousevolution} in comparison to the actual mean evolution. \label{fig:mean_perturbation_evolution}}
\end{figure}

We empirically found agreement with the dynamics predicted in Eq.~\eqref{eq:meanevolution} as was shown in Fig.~\ref{fig:inhomogeneousevolution}. To ensure consistency with the evolution in the homogeneous and isotropic case we also check the predictions for the mean field using the mean value $\alpha$ of the NTK instead of  using $\lambda_1/N.$ The resulting discrepancy is shown in Figure~\ref{fig:mean_perturbation_evolution}. We find no significant deviations between the two predictions across our network sample.

\bibliographystyle{inspire}
\bibliography{references2}

\providecommand{\href}[2]{#2}\begingroup\raggedright\begin{thebibliography}{10}

\bibitem{polyak}
B.~Polyak, ``Some methods of speeding up the convergence of iteration
  methods,'' \href{http://dx.doi.org/10.1016/0041-5553(64)90137-5}{Ussr
  Computational Mathematics and Mathematical Physics {\bfseries 4} (12, 1964)
  1--17}.

\bibitem{Mukhanov:2005sc}
V.~Mukhanov, {\em {Physical Foundations of Cosmology}}.
\newblock Cambridge University Press, Oxford, 2005.

\bibitem{Weinberg:2008zzc}
S.~Weinberg, {\em {Cosmology}}.
\newblock 2008.

\bibitem{Baumann:2009ds}
D.~Baumann,
  \href{http://dx.doi.org/10.1142/9789814327183_0010}{``{Inflation},''} in {\em
  {Theoretical Advanced Study Institute in Elementary Particle Physics}:
  {Physics of the Large and the Small}}.
\newblock 7, 2009.
\newblock \href{http://arxiv.org/abs/0907.5424}{[arXiv:0907.5424 [hep-th]]}.

\bibitem{ParticleDataGroup:2020ssz}
{\bfseries Particle Data Group} Collaboration, P.~A. Zyla {\em et~al.},
  ``{Review of Particle Physics},''
  \href{http://dx.doi.org/10.1093/ptep/ptaa104}{PTEP {\bfseries 2020} no.~8,
  (2020) 083C01}.

\bibitem{ntk1}
A.~Jacot, F.~Gabriel, and C.~Hongler, ``Neural tangent kernel: Convergence and
  generalization in neural networks,'' CoRR {\bfseries abs/1806.07572} (2018) ,
  \href{http://arxiv.org/abs/1806.07572}{[1806.07572]}.

\bibitem{1902.06720}
J.~Lee, L.~Xiao, S.~S. Schoenholz, Y.~Bahri, R.~Novak, J.~Sohl-Dickstein, and
  J.~Pennington, ``Wide neural networks of any depth evolve as linear models
  under gradient descent,''
  \href{http://dx.doi.org/10.1088/1742-5468/abc62b}{Journal of Statistical
  Mechanics: Theory and Experiment {\bfseries 2020} no.~12, (Dec, 2020)
  124002}, \href{http://arxiv.org/abs/1902.06720}{[arXiv:1902.06720
  [stat.ML]]}.

\bibitem{novak2019neural}
R.~Novak, L.~Xiao, J.~Hron, J.~Lee, A.~A. Alemi, J.~Sohl-Dickstein, and S.~S.
  Schoenholz, ``Neural tangents: Fast and easy infinite neural networks in
  python,'' arXiv preprint arXiv:1912.02803 (2019) .

\bibitem{Linde:1983gd}
A.~D. Linde, ``{Chaotic Inflation},''
  \href{http://dx.doi.org/10.1016/0370-2693(83)90837-7}{Phys. Lett. B
  {\bfseries 129} (1983) 177--181}.

\bibitem{Qian}
N.~Qian, ``On the momentum term in gradient descent learning algorithms,''
  Neural networks {\bfseries 12} no.~1, (1999) 145--151.

\bibitem{Su}
W.~Su, S.~Boyd, and E.~J. Candes, ``A differential equation for modeling
  nesterov's accelerated gradient method: Theory and insights,'' The Journal of
  Machine Learning Research {\bfseries 17} no.~1, (2016) 5312--5354.

\bibitem{Roberts:2021fes}
D.~A. Roberts, S.~Yaida, and B.~Hanin, ``{The Principles of Deep Learning
  Theory},'' \href{http://arxiv.org/abs/2106.10165}{[arXiv:2106.10165
  [cs.LG]]}.

\bibitem{neal2012bayesian}
R.~M. Neal, {\em Bayesian learning for neural networks}, vol.~118.
\newblock Springer Science \& Business Media, 2012.

\bibitem{lee2017deep}
J.~Lee, Y.~Bahri, R.~Novak, S.~S. Schoenholz, J.~Pennington, and
  J.~Sohl-Dickstein, ``Deep neural networks as gaussian processes,'' arXiv
  preprint arXiv:1711.00165 (2017) .

\bibitem{matthews2018gaussian}
A.~G. d.~G. Matthews, M.~Rowland, J.~Hron, R.~E. Turner, and Z.~Ghahramani,
  ``Gaussian process behaviour in wide deep neural networks,'' arXiv preprint
  arXiv:1804.11271 (2018) .

\bibitem{novak2018bayesian}
R.~Novak, L.~Xiao, J.~Lee, Y.~Bahri, G.~Yang, J.~Hron, D.~A. Abolafia,
  J.~Pennington, and J.~Sohl-Dickstein, ``Bayesian deep convolutional networks
  with many channels are gaussian processes,'' arXiv preprint arXiv:1810.05148
  (2018) .

\bibitem{garriga2018deep}
A.~Garriga-Alonso, C.~E. Rasmussen, and L.~Aitchison, ``Deep convolutional
  networks as shallow gaussian processes,'' arXiv preprint arXiv:1808.05587
  (2018) .

\bibitem{Halverson:2020trp}
J.~Halverson, A.~Maiti, and K.~Stoner, ``{Neural Networks and Quantum Field
  Theory},'' \href{http://dx.doi.org/10.1088/2632-2153/abeca3}{Mach. Learn.
  Sci. Tech. {\bfseries 2} no.~3, (2021) 035002},
  \href{http://arxiv.org/abs/2008.08601}{[arXiv:2008.08601 [cs.LG]]}.

\bibitem{Maiti:2021fpy}
A.~Maiti, K.~Stoner, and J.~Halverson, ``{Symmetry-via-Duality: Invariant
  Neural Network Densities from Parameter-Space Correlators},''
  \href{http://arxiv.org/abs/2106.00694}{[arXiv:2106.00694 [cs.LG]]}.

\bibitem{Erbin:2021kqf}
H.~Erbin, V.~Lahoche, and D.~O. Samary, ``{Nonperturbative renormalization for
  the neural network-QFT correspondence},''
  \href{http://arxiv.org/abs/2108.01403}{[arXiv:2108.01403 [hep-th]]}.

\bibitem{Halverson:2021aot}
J.~Halverson, ``{Building Quantum Field Theories Out of Neurons},''
  \href{http://arxiv.org/abs/2112.04527}{[arXiv:2112.04527 [hep-th]]}.

\bibitem{Luo:2021nab}
D.~Luo and J.~Halverson, ``{Infinite Neural Network Quantum States},''
  \href{http://arxiv.org/abs/2112.00723}{[arXiv:2112.00723 [quant-ph]]}.

\bibitem{Dyer:2019uzd}
E.~Dyer and G.~Gur-Ari, ``{Asymptotics of Wide Networks from Feynman
  Diagrams},'' \href{http://arxiv.org/abs/1909.11304}{[arXiv:1909.11304
  [cs.LG]]}.

\bibitem{yang2021feature}
G.~Yang and E.~J. Hu, ``Feature learning in infinite-width neural networks,''
  2021.

\bibitem{yang2021tensor}
G.~Yang, ``Tensor programs i: Wide feedforward or recurrent neural networks of
  any architecture are gaussian processes,'' 2021.

\bibitem{yang2020scaling}
G.~Yang, ``Scaling limits of wide neural networks with weight sharing: Gaussian
  process behavior, gradient independence, and neural tangent kernel
  derivation,'' 2020.

\bibitem{yang2020tensor}
G.~Yang, ``Tensor programs ii: Neural tangent kernel for any architecture,''
  2020.

\bibitem{yang2021tensor3}
G.~Yang, ``Tensor programs iii: Neural matrix laws,'' arXiv preprint
  arXiv:2009.10685 (2020) .

\bibitem{Mukhanov:1990me}
V.~F. Mukhanov, H.~A. Feldman, and R.~H. Brandenberger, ``{Theory of
  cosmological perturbations. Part 1. Classical perturbations. Part 2. Quantum
  theory of perturbations. Part 3. Extensions},''
  \href{http://dx.doi.org/10.1016/0370-1573(92)90044-Z}{Phys. Rept. {\bfseries
  215} (1992) 203--333}.

\bibitem{DeLuca:2022brp}
G.~B. De~Luca and E.~Silverstein, ``{Born-Infeld (BI) for AI: Energy-Conserving
  Descent (ECD) for Optimization},''
  \href{http://arxiv.org/abs/2201.11137}{[arXiv:2201.11137 [cs.LG]]}.

\bibitem{Barcelo:2005fc}
C.~Barcelo, S.~Liberati, and M.~Visser, ``{Analogue gravity},''
  \href{http://dx.doi.org/10.12942/lrr-2005-12}{Living Rev. Rel. {\bfseries 8}
  (2005) 12}, \href{http://arxiv.org/abs/gr-qc/0505065}{[arXiv:gr-qc/0505065]}.

\bibitem{Gielen:2013kla}
S.~Gielen, D.~Oriti, and L.~Sindoni, ``{Cosmology from Group Field Theory
  Formalism for Quantum Gravity},''
  \href{http://dx.doi.org/10.1103/PhysRevLett.111.031301}{Phys. Rev. Lett.
  {\bfseries 111} no.~3, (2013) 031301},
  \href{http://arxiv.org/abs/1303.3576}{[arXiv:1303.3576 [gr-qc]]}.

\bibitem{Maldacena:1997re}
J.~M. Maldacena, ``{The Large N limit of superconformal field theories and
  supergravity},'' \href{http://dx.doi.org/10.1023/A:1026654312961}{Adv. Theor.
  Math. Phys. {\bfseries 2} (1998) 231--252},
  \href{http://arxiv.org/abs/hep-th/9711200}{[arXiv:hep-th/9711200]}.

\bibitem{Verlinde:2010hp}
E.~P. Verlinde, ``{On the Origin of Gravity and the Laws of Newton},''
  \href{http://dx.doi.org/10.1007/JHEP04(2011)029}{JHEP {\bfseries 04} (2011)
  029}, \href{http://arxiv.org/abs/1001.0785}{[arXiv:1001.0785 [hep-th]]}.

\bibitem{Burgess:2013ara}
C.~P. Burgess,
  \href{http://dx.doi.org/10.1093/acprof:oso/9780198728856.003.0004}{``{The
  Cosmological Constant Problem: Why it's hard to get Dark Energy from
  Micro-physics},''} in {\em {100e Ecole d'Ete de Physique: Post-Planck
  Cosmology}}, pp.~149--197.
\newblock 2015.
\newblock \href{http://arxiv.org/abs/1309.4133}{[arXiv:1309.4133 [hep-th]]}.

\bibitem{gadget}
V.~{Springel}, N.~{Yoshida}, and S.~D.~M. {White}, ``{GADGET: a code for
  collisionless and gasdynamical cosmological simulations},''
  \href{http://arxiv.org/abs/astro-ph/0003162}{[arXiv:astro-ph/0003162
  [astro-ph]]}.

\bibitem{Felder:2000hq}
G.~N. Felder and I.~Tkachev, ``{LATTICEEASY: A Program for lattice simulations
  of scalar fields in an expanding universe},''
  \href{http://dx.doi.org/10.1016/j.cpc.2008.02.009}{Comput. Phys. Commun.
  {\bfseries 178} (2008) 929--932},
  \href{http://arxiv.org/abs/hep-ph/0011159}{[arXiv:hep-ph/0011159]}.

\end{thebibliography}\endgroup
\end{document}